\documentclass{elsart}
\usepackage{graphicx}

\begin{document}

\def\Beqa{\begin{eqnarray}}
\def\Eeqa{\end{eqnarray}}
\def\Bq{{\bf q}}

\newcommand{\BSIG}{\mbox{\boldmath $\sigma$}}

\def\lsim{\mathrel{\mathpalette\gl@align<}}
\def\gsim{\mathrel{\mathpalette\gl@align>}}

\begin{frontmatter}

\title{Study on the Kondo effect in the tunneling phenomena 
through a quantum dot}
\author[address1]{Osamu Sakai\thanksref{thank1}},
\author[address2]{Wataru Izumida}

\address[address1]{Department of Physics, Tokyo Metropolitan University, Tokyo
192-0397, Japan}
\address[address2]{Mesoscopic Correlation Project, ERATO, JST, NTT
 Atsugi R \& D Center, Atsugi 243-0198, Japan}

\thanks[thank1]{Corresponding author.
 Department of Physics, Tokyo Metropolitan University, Tokyo
 192-0397, Japan
 E-mail: sakai@phys.metro-u.ac.jp}

\begin{abstract}
We review our recent studies on the Kondo effect in the tunneling phenomena 
through quantum dot systems.
Numerical methods to calculate reliable 
tunneling conductance 
are developed.
In the first place, a case in which electrons of odd number occupy
the dot is  studied,  and 
experimental results are analyzed based on the calculated result.
Tunneling anomaly in the even-number-electron occupation case, 
which is recently 
observed in experiment and is ascribed to the Kondo effect in the 
spin singlet-triplet cross over transition region, 
is also examined theoretically.
\end{abstract}

\end{frontmatter}

\section{Introduction}
Quantum dot systems are now designed as the artificial magnetic
impurity and are growing as a field of detailed experimental
studies of the Kondo problems\cite{A0} . 
In this report we review our recent theoretical works on the Kondo
effect in the tunneling phenomena through  quantum 
dot\cite{A1,A2,A3}.

After the papers pointed out the possibility to occur the Kondo effect 
in tunneling through a quantum dot\cite{B0}, many theoretical studies 
have been done on this problem\cite{C0}.
The calculation of the tunneling conductance needs the dynamical
excitation  spectra.
However we have not exact analytic calculation of the dynamical 
excitation spectra for the Kondo systems\cite{A4}.
Two numerical methods have been recently developed to calculate the 
tunneling conductance of the quantum dot systems.
One is based on the numerical renormalization group technique
(NRG)\cite{A5}, and 
another is based on the Quantum Monte Carlo (QMC) technique\cite{A2}.
Both techniques are known as reliable  methods to calculate the dynamical
excitation of the Kondo systems\cite{A4,A6,A7}.

When the occupation number of electrons on the dot is odd,
localized spin freedom appears on the dot, and it couples with the 
conduction electrons on the leads.
At very low temperatures,
we can expect the increase of the tunneling conductance 
due to the 
resonance transmission  {\it via} the Kondo  peak in the  density of 
states on the dot orbitals.
This is the most typical example of the Kondo effect of the dot 
systems, and has been observed in many experiments\cite{A0}.
In the first part of this report, we present the theoretical 
calculation for this case\cite{A1}
and compare it with experimental data\cite{A8}.
Recently, anomaly in a region of an even electron number occupation 
case has been reported in experiment\cite{A9}.
This phenomena is expected to relate to the Kondo effect in the 
spin crossover region of even occupation number case\cite{A3,A10,A11}.
This problem is discussed in the second part of this report.

\section{Single Orbital Case}

We consider the following Hamiltonian,
\begin{eqnarray}
H & = & H_{\rm \ell}+H_{\rm d}+H_{\rm \ell-d}, \\
H_{\rm \ell}& = & \sum_{\mu k\sigma}
\varepsilon_{\mu k} c_{\mu k\sigma}^{+}c_{\mu k\sigma}, \\
H_{\rm d}& = & \sum_{p\sigma}\varepsilon_{dp}n_{dp\sigma}
+U\sum_{<p\sigma,p'\sigma>}n_{dp\sigma}n_{dp'\sigma'}, \\ 
H_{\rm \ell-d}& = & \frac{1}{\sqrt{N}}\sum_{p\mu k\sigma}
\{v_{p\mu k}d_{p\sigma}^{+}c_{\mu k\sigma}+
{\rm h.c.}\}.
\end{eqnarray}
The terms $H_{\rm \ell}$ and $H_{\rm d}$ represent the electron in the 
leads and the dot, respectively. 
The term $H_{\rm \ell-d}$ gives the electron tunneling between the leads 
and the dot. 
The suffix $\mu=L(R)$ means the left(right) lead and $dp$ means the dot 
orbital denoted by $p$.
The quantity $\varepsilon_{dp}$ corresponds to the energy of the orbital, 
and it can be changed by applying gate voltage. 
The quantity $U$ is the Coulomb interaction constant.

At first we consider the most simplified model that the dot has 
a single orbital. 
We abbreviate the suffix $p$.
There will be many orbitals in dot in actual situations. 
Two orbitals case will be discussed in \S 3.
We calculate the conductance, $G$, in the linear theory of the bias voltage.
It is obtained from the correlation 
function of the current operators\cite{A5}.
But in the single orbital case, the calculation is reduced 
to the following expression\cite{A2},
\begin{equation}
G=\frac{2 e^{2}}{h}\int\frac{4\Gamma_{L}\Gamma_{R}}{\Gamma_{L}+\Gamma_{R}}
(-{\rm Im}G_{dd}(\epsilon))
\bigl(-\frac{\partial f}{\partial \epsilon}\bigr){\rm d}\epsilon,
\label{E3.1}
\end{equation}
where $G_{dd}(\epsilon)$ is Green's function of the dot orbital and 
$f(\epsilon)$ is the Fermi distribution function.
We use this formula in this section because the numerical calculation of 
$G_{dd}(\epsilon)$ is easier than the calculation of the current correlation
function.
Detailed comparison of both methods are made in Ref.\cite{A2}.
Hereafter we assume that the leads have constant density of states 
from $-D$ to $D$ with $D=1$.
The hybridization strength is parameterized as 
$\Gamma_{\mu}=\pi |v_{\mu}|^{2}\rho_{c}$, where $\rho_{c}=1/2D$ 
is the density 
of states for the lead.

\begin{figure}[btp]
%h=here, t=top, b=bottom, p=separate figure page
\begin{center}\leavevmode
\includegraphics[width=0.8\linewidth]{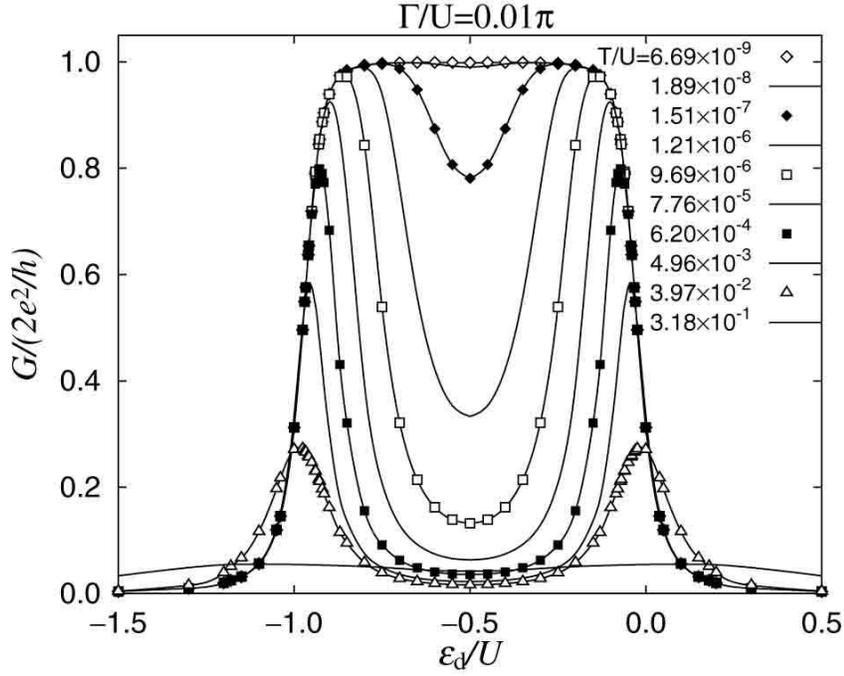}
%%%Fig.1
%%%\vspace{5cm}
\caption{ 
Conductance as a function of $\varepsilon_{d}$ for the parameter
$\Gamma/(\pi U)=1.0\times 10^{-2}$ 
($ U= 1.0 \times 10^{-2},\Gamma/\pi=1.0 \times 10^{-4}$).
}\label{F1}\end{center}\end{figure}

In Fig.\ref{F1}, we show the conductance as a function of the gate voltage 
($\varepsilon_{d}$) for various temperature cases\cite{A1}.
This calculation is carried out by the NRG method.
At higher temperatures we have paired Coulomb oscillation peaks at 
$\varepsilon_{d}=-U$ and 0.
They grow without increase of their width, thus become very sharp as the 
temperature decreases.
At the same time the peak positions shift slightly to $\varepsilon_{d}=-U/2$
side.
When the temperature decreases further, 
the intensity  of the  valley region between the two peaks 
gradually increases, 
and the peaks merge into a broad single peak at extremely low temperature.
The characteristic temperature $T_{\rm KM}$ 
varies drastically as $\varepsilon_{d}$ changes.
(We define $T_{\rm KM}$ 
following the usual definition, 
$T_{\rm K}=4\chi(T=0)$,
for each 
$\varepsilon_{d}$ case).
It is the lowest at the mid point of the two peaks,
$\varepsilon_{d}=-U/2$,
and is denoted as $T_{\rm KM}^{*}$, which is 
$T_{\rm KM}^{*}/U=5.34\times10^{-7}$ for the parameter case used in 
Fig.\ref{F1}. 
The conductance at the mid point begin to increase at about 
$10T_{\rm KM}^{*}$, and the valley disappears at about $0.2T_{\rm KM}^{*}$.

\begin{figure}[btp]
%h=here, t=top, b=bottom, p=separate figure page
\begin{center}\leavevmode
\includegraphics[width=0.8\linewidth]{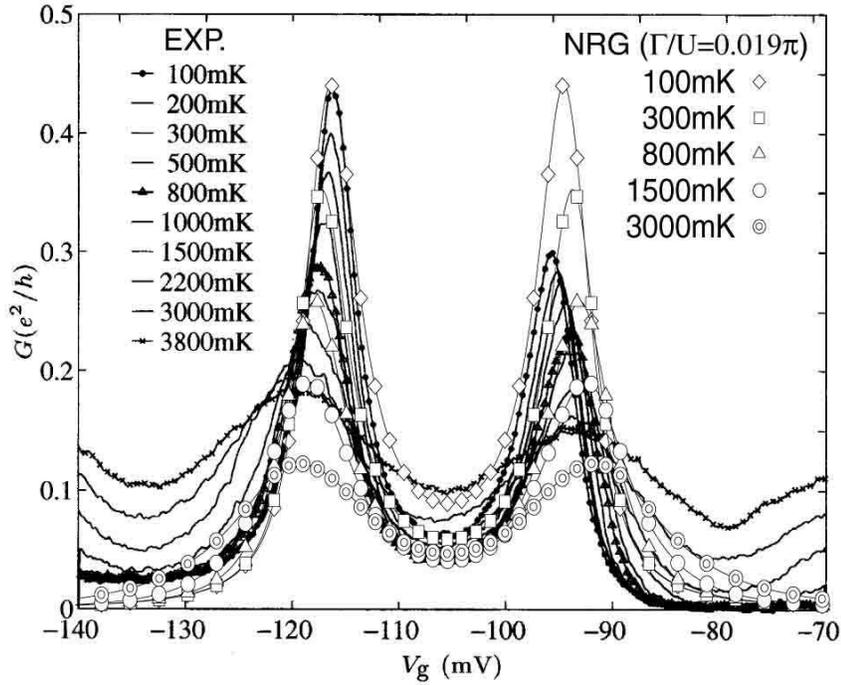}
%%%Fig.2
%%%\vspace{5cm}
\caption{ 
Comparison between the experimental and the calculated data.
The experimental data is reproduced from Fig.2 of Ref.\cite{A8}.
The parameter of the NRG calculation is chosen to be 
$\Gamma/(\pi U)=1.9 \times 10^{-2}$
}\label{F2}\end{center}\end{figure}

Goldhaber-Gordon et al. have shown the detailed temperature 
dependence of the conductance\cite{A8}.
We directly compare the NRG data with experimental data 
in Fig.\ref{F2}\cite{A1}.
The Coulomb repulsion has been estimated to be 1.9 meV.
We have estimated $\Gamma = 0.12$meV =($1.3\times 10^{3}$mK), 
and thus $\Gamma/\pi U = 1.9 \times 10^{-2}$ in the experimental 
situation.
On horizontal axis the points $\varepsilon_{d}=0, -U$ of the calculated
conductance are set at $V_{g} = -119, -92$meV, respectively, and factor
0.31 is multiplied to the NRG data to fit the experimental data at 
the lowest temperature.
This factor 0.31 would be caused by the asymmetry 
$\Gamma_{\rm L} \neq \Gamma_{\rm R}$.
For the left hand side peak, the conductance data agrees very well with 
the experimental one in 100 mK $\sim$ T $\sim$ 1500mK.
This agreement suggests  that the behaviors in experiment are caused
by the Kondo effect in the mid-temperature region shown in Fig.\ref{F1}.
The Kondo temperature at the valley region seems to be less than 10 mK.

There are several discrepant points. At high temperature region 
T $\ge$ 2000mK, the conductance of the experimental data is larger 
than that of the NRG data.
This might be caused by the multi-orbital effect.
The conductance shows disagreement in the valley and the 
right hand peak positions.
The change of the gate voltage on the dot will affect not only the 
potential of the dot, but also the hybridization strength between the 
dot and the lead states.

The conductance at low temperature is strongly suppressed by application
of the Zeeman field.
The theoretical study has been carried out based on the QMC
method\cite{A2},
and the result agrees well with experimental one\cite{B1}.

\section{Spin Crossover Transition}
Recently, Sasaki et al. observed low temperature tunneling anomaly in 
even-number-electron-occupation case\cite{A9}.
They controlled the energy splitting of orbitals by tuning the magnetic
field\cite{B2}.
When the level splitting is gradually increased in the even electron
number case,
the energy of the spin triplet state
increases compared with that of the singlet state.
The electrons occupy different orbitals to gain Hund's coupling 
energy in the former state,
while  the electrons with opposite spin occupy the lower energy orbital
in the latter state. 
It was suggested that 
the low temperature anomaly is related to the Kondo effect due 
to this spin crossover transition\cite{A10,A11}.
However detailed calculation of the conductance has not been 
done.
It is observed that a bump grows between the two Coulomb peaks,
i.e., a third peak appears between the Coulomb peaks and it grows 
when the temperature decreases.
This behavior is quite different from that of the odd number case
discussed in previous section.

We calculate the conductance around the singlet-triplet crossover 
region for a system with two orbitals\cite{A3}.
The orbitals are denoted as even($p$=e) and odd($p$=o). 
The energy of e(o) orbital is defined as 
$\varepsilon_{d}-\delta/2$
($\varepsilon_{d}+\delta/2$).
The exchange term
$
H_{\rm ex}
=J_{\rm H}\sum_{\sigma_{1}\sigma_{2}\sigma_{3}\sigma_{3}}
(\BSIG)_{\sigma_{1}\sigma_{2}}
(\BSIG)_{\sigma_{3}\sigma_{4}}
d^{+}_{{\rm e}\sigma_{1}}
d_{{\rm e}\sigma_{2}}
d^{+}_{{\rm o}\sigma_{3}}
d_{{\rm o}\sigma_{4}}
$
is added to the Hamiltonian Eq.(1). ($J_{\rm H} < 0$)

\begin{figure}[btp]
%h=here, t=top, b=bottom, p=separate figure page
\begin{center}\leavevmode
\includegraphics[width=0.8\linewidth]{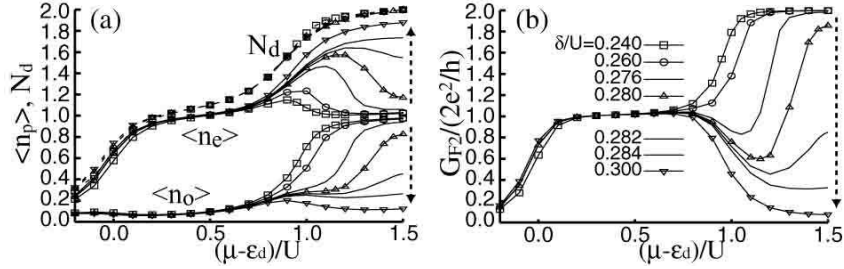}
%%%Fig.3
%%%\vspace{5cm}
\caption{ 
(a) Occupation numbers of the e and o orbitals at $T=0$,
$\langle n_{\rm e} \rangle$, 
$\langle n_{\rm o} \rangle$, 
and the total occupation number on the dot, $N_{\rm d}$.
(b) Conductance at $T=0$ for the two channel case, $G_{F2}$.
The symbols for $\delta$ are common in (a) and (b).
The dashed arrows show the increasing of $\delta$. 
Parameters are $J_{\rm H}/{U}=-0.3$,
$\Gamma_{\rm e}/U=\Gamma_{\rm o}/U
=0.02$.
}\label{F5}\end{center}\end{figure}

\begin{figure}[btp]
%h=here, t=top, b=bottom, p=separate figure page
\begin{center}\leavevmode
\includegraphics[width=0.8\linewidth]{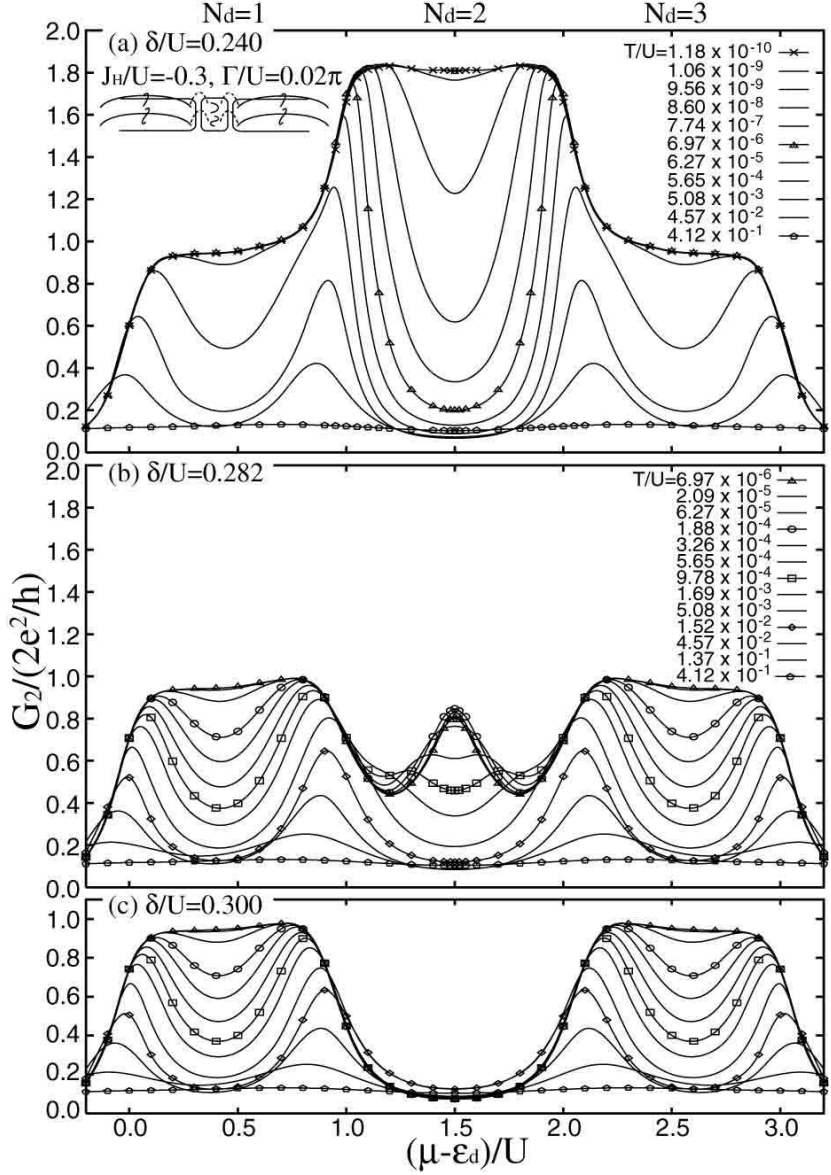}
%%%Fig.4
%%%\vspace{5cm}
\caption{ 
Conductance at various temperatures as a function of gate voltage 
in the two channel case.
(a)$\delta/U=0.24$,
(b)$\delta/U=0.282$,
(c)$\delta/U=0.3$.
The symbols of the temperature in (b) and (c) are common.
}\label{F6}\end{center}\end{figure}

\begin{figure}[btp]
%h=here, t=top, b=bottom, p=separate figure page
\begin{center}\leavevmode
\includegraphics[width=0.8\linewidth]{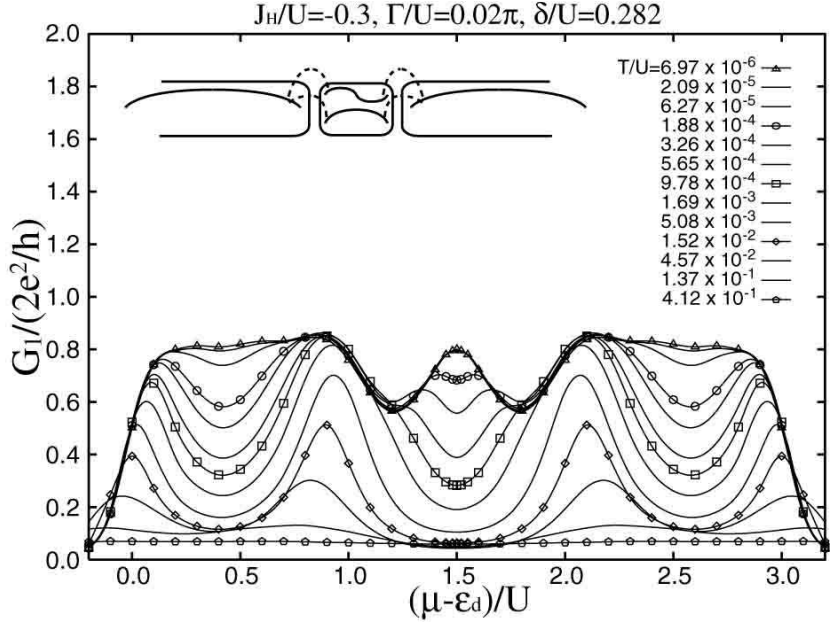}
%%%Fig.5
%%%\vspace{5cm}
\caption{ 
Conductance as a function of gate voltage in the single channel case.
Parameters are the same to those of 
Fig.\ref{F6}(b).
}\label{F7}\end{center}\end{figure}

In Fig.\ref{F5}(a), we show the occupation numbers of the even and 
odd orbitals, 
$\langle n_{\rm e} \rangle$ and $\langle n_{\rm o} \rangle$, on the dot.
The parameters are $J_{\rm H}/U=-0.3$, $\Gamma_{\rm e}/U=\Gamma_{\rm
o}/U=0.02\pi$.

We assume the case in which 
both leads have two conduction channels, and each dot orbital connects
to each conduction channel (two channel case).
The conductance is given as the sum of ones from two channels.
At very low temperatures it is given as,  
$
G_{\rm F2} = (2e^{2}/h) \sum_{p}\sin^{2} (\pi \langle n_{p} \rangle /2)
$,
and is shown in Fig.\ref{F5}(b).

Here we  define a quantity $x=(\mu-\varepsilon_{d})/U$.
In the $0 \sim x \sim 1$ region, an electron occupies the e-orbital, 
$(\langle n_{\rm e} \rangle, \langle n_{\rm o} \rangle) \simeq (1,0)$.
Next we see the $1 \sim x \sim 1.5$ region.
At $\delta/U=0.24$, the occupations on both orbitals are almost the
same, $(\langle n_{\rm e} \rangle, \langle n_{\rm o} \rangle) \simeq
(1,1)$, due to the strong Hund's coupling.
When $\delta$ increases, the occupations in the $1 \sim x \sim 1.5$
 region gradually 
split to $(2,0)$ where the local spin state is a singlet.
Therefore the conductance decreases from $4e^{2}/h$ to $0$.
We stress that the splitting is suppressed around $x=1.5$ as seen in 
Fig.\ref{F5}(a).
For example at $\delta/U=0.28$, the second electron tends to occupy the e-orbital when $x$ sweeps to $x=1.2$ from smaller $x$, then they {\it redistribute} to the e- and o-orbitals when $x$ further sweeps to $x=1.5$; $(1,0) \rightarrow (1.6,0.3) \rightarrow (1.2,0.8)$.
This redistribution reflects the stronger Hund's coupling near $x=1.5$, at which the total occupation is close to two.
The redistribution causes a bump in the conductance around $x=1.5$ at low temperatures, as seen in $0.276 \sim \delta/U \sim 0.284$ in Fig.\ref{F5}(b).

In Fig.\ref{F6} we show the conductance at various temperatures 
around the local spin singlet-triplet degeneracy.
The parameters are: (a) $\delta/U=0.24$, (b), $\delta/U=0.282$, 
(c) $\delta/U=0.3$.
There are the four Coulomb peaks on $x = (\mu-\varepsilon_{\rm d})/U \sim 0$, $1$, $2$, $3$ at high temperatures.
When the temperature decreases, the conductance in the region $0 \sim x \sim 1$ ($2 \sim x \sim 3$), where $N_{\rm d} \simeq 1$ ($N_{\rm d} \simeq 3$), increases to $\simeq 2e^{2}/h$, caused by the usual spin-1/2 Kondo effect for the e-channel (o-channel).
On the other hand,  the region $1 \sim x \sim 2$, 
where $N_{\rm d} \simeq 2$, is remarkable.
At (a) $\delta/U=0.24$, the conductance is large, about $4e^{2}/h$.
However, the large conductance appears at extremely low temperature (e.g., $T/U=10^{-8}$ corresponds to $10^{-3}{\rm mK}$ for $U\simeq{\rm 10meV}$).
When $\delta$ increases to (b) $\delta/U=0.282$, a "bump" emerges in the region $1.2 \sim x \sim 1.8$ as the temperature decreases.
For the case (c) $\delta/U=0.3$, the conductance in the region 
$1 \sim x \sim 2$ is nearly constant with a small value.
The behaviors (a)-(c) are classified into, 
(a) the local spin triplet Kondo effect, 
(b) the singlet-triplet Kondo effect, and 
(c) the usual even-odd oscillations for the dot with large energy 
separation~\cite{A0,A5}.

Next we consider a case in which leads have only one conduction
channel,
and the both dot orbitals connect to this single channel (single channel
case).
In such case one orbital will hybridize mainly with even combination 
of L and R lead states, and another mainly with odd combination
of L and R states.
The tunneling {\it via} different orbitals interferes in this case.
At $T=0$, the conductance is given as
$
G_{\rm F1} = (2e^{2}/h)\sin^{2} (\pi( 
\langle n_{e} \rangle -\langle n_{o} \rangle)
/2)
$
, and is  small when 
$
\langle n_{e} \rangle \simeq \langle n_{o} \rangle
$.
Therefore the conductance will tend to zero as $T$ decreases to extreme
low temperature when Hund's rule coupling energy dominates.
This is contrasted to the case (a) in Fig.\ref{F6}.

In the singlet-triplet Kondo effect case  where the Kondo 
temperature is not low,
the calculated temperature dependence 
of the conductance is not so different from that of the two channel 
case(Fig.\ref{F6}(b)) as seen from Fig.\ref{F7}.
This is because
$\langle n_{e} \rangle$ and $\langle n_{o} \rangle$ have different and 
not-integer values as seen from Fig.\ref{F5}(a).

As a summary, we have shown that the electron occupation on orbitals 
redistribute to gain Hund's coupling energy
in the singlet-triplet crossover region when the potential deepens.
This redistribution causes a bump, as seen in the experiment, 
 in the conductance at low temperatures.

\begin{ack}
The authors would like to thank  Professor Tarucha for collaboration.
The numerical computation was partly performed in the Computer Center
of Tohoku University and the Supercomputer Center of Institute for Solid
State Physics (University of Tokyo).
\end{ack}


\begin{thebibliography}{9}
\bibitem{B2}
We can neglect the Zeeman splitting in this case, see for example \cite{A3}.
\end{thebibliography}
\end{document}